\documentclass[letterpaper,journal]{IEEEtran}
\usepackage{amsmath,amsfonts}
\usepackage{algorithmic}
\usepackage{algorithm}
\usepackage{array}
\usepackage[caption=false,font=normalsize,labelfont=sf,textfont=sf]{subfig}
\usepackage{textcomp}
\usepackage{stfloats}
\usepackage{url}
\usepackage{verbatim}
\usepackage{graphicx}
\usepackage{cite}
\usepackage{booktabs}
\usepackage{multirow}
\usepackage[table]{xcolor}
\definecolor{resultbg}{RGB}{226,239,248}
\newcommand{\ourbest}[1]{\cellcolor{resultbg}\textbf{#1}}
\usepackage[hidelinks]{hyperref}
\begin{document}

\title{DualAffect: Dual-Path Paralinguistic Integration for Generative Audiovisual Emotion Understanding}
\author{%
Yibo~Wang,
Jisheng~Dang,
Bimei~Wang,
Junhong~Liu,
Jiongpei~Hou,
Jingwen~Zhao,
Wenjun~Ma,
Hong~Peng,
Bin~Hu,~\IEEEmembership{Fellow,~IEEE,}
Qi~Tian,~\IEEEmembership{Fellow,~IEEE,}
and~Tat-Seng~Chua%
\thanks{Yibo Wang, Jisheng Dang, Bimei Wang, Junhong Liu, Jiongpei Hou, Jingwen Zhao, Wenjun Ma, Hong Peng, and Bin Hu are with Lanzhou University, Lanzhou, China.}%
\thanks{Qi Tian is with Cloud and AI BU, Huawei, Shenzhen, Guangdong 518129, China (e-mail: tian.qi1@huawei.com).}%
\thanks{Tat-Seng Chua is with the School of Computing, National University of Singapore, Singapore.}%
\thanks{Corresponding authors: Jisheng Dang, Bimei Wang, and Bin Hu.}%
}
\maketitle

\begin{abstract}
Understanding human emotion is essential for multimodal systems that interpret both the content of an interaction and the speaker's expressive state.
However, connecting paralinguistic cues to generated answers remains challenging, particularly when vocal expression diverges from visual appearance.
We propose DualAffect, a dual-path architecture that integrates visually conditioned acoustic features and temporally organized acoustic tokens into a shared language decoder.
The two pathways reuse the same acoustic encoding.
The fusion pathway associates sound with visual content through visual queries, while the direct pathway compresses the valid acoustic sequence into ordered tokens through an independent projector.
This design gives the decoder access to both joint audiovisual context and acoustic evidence organized along its own timeline.
Mixed-task adaptation with task-dependent output preservation supports emotion recognition and general audiovisual question answering within the same model.
DualAffect achieves an average metric of 67.89 on CA-MER and accuracies of 62.88\% on the EmotionTalk test set and 49.87\% on Daily-Omni.
Pathway interventions show that the trained decoder uses direct acoustic evidence alongside existing fusion.
With model parameters and fusion inputs fixed, zeroing the direct tokens reduces EmotionTalk validation accuracy by 9.75 percentage points and the CA-MER average metric by 2.66 points.
The largest CA-MER subset effect occurs when the overall emotion agrees with audio but differs from vision.
These findings identify acoustic representation organization as a practical design axis for connecting paralinguistic perception to unified audiovisual generation.
Code is available at \url{https://github.com/YiboWang2002/DualAffect}.
\end{abstract}


\section{Introduction}
\label{sec:introduction}

\suppressfloats[t]
\begin{figure}[t]
  \centering
  \includegraphics[width=\columnwidth,pagebox=cropbox]{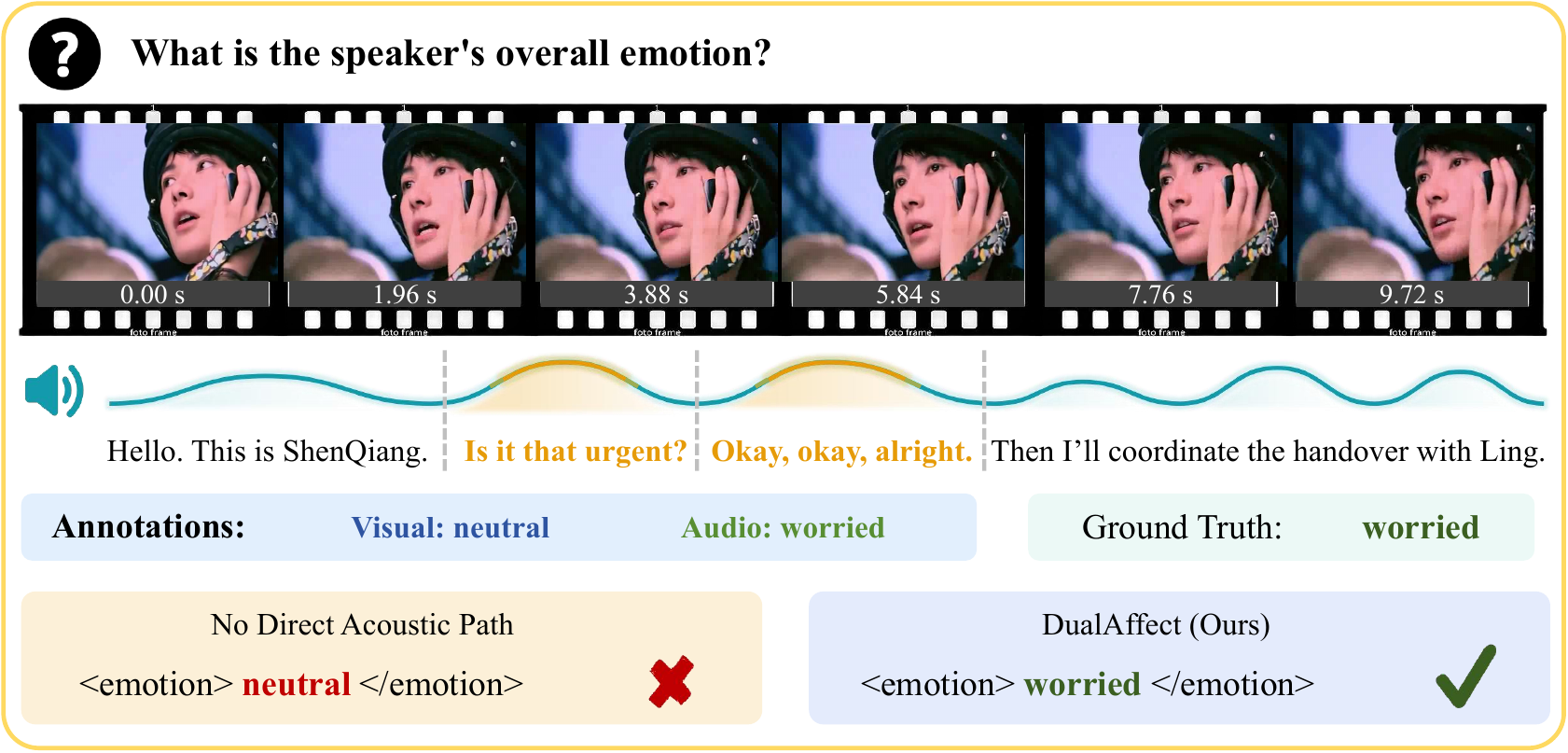}
  \caption{Direct acoustic access in an audio-aligned example. The visual label is neutral, while the audio and overall labels are worried. Zeroing the direct tokens changes DualAffect's answer from worried to neutral with audiovisual fusion retained. The curve above the translated dialogue is a schematic illustration of vocal prosody.}
  \label{fig:motivation}
\end{figure}

\IEEEPARstart{U}{nderstanding} human emotion requires interpreting both what a person says and how they say it.
Intonation, rhythm, and voice quality convey emotion even when vocal delivery diverges from facial expressions or verbal content.
Multimodal language models such as Emotion-LLaMA and AffectGPT extend recognition to emotion description through affective instruction tuning\cite{introCheng2024EmotionLLaMA,introLian2025AffectGPT}.
This shift requires integrating affective evidence into answer generation to interpret both interaction content and the speaker's expressive state.

Effective affective generation depends on whether encoded acoustic cues influence the answer.
LISTEN reports substantial lexical dependence when audio language models must infer emotion from vocal delivery\cite{introChen2026LISTEN}.
VoxParadox further identifies a gap between paralinguistic information recoverable from internal representations and its use in generated responses\cite{introPang2026VoxParadox}.
These findings motivate studying the connection between acoustic representations and generative decisions.
In audiovisual understanding, this connection involves two reference structures, the visual context in which speech occurs and the temporal sequence along which vocal expression unfolds.
We ask how both structures can jointly inform emotion answers. Fig.~\ref{fig:motivation} illustrates a case in which direct acoustic access changes the answer while audiovisual fusion remains available.

Prior work advances affective generation through complementary encoders, cross-modal interaction, and targeted supervision.
WavLLM combines speech encoders, while multitask affective reasoning incorporates emotion-specific representations\cite{introHu2024WavLLM,introZhang2025BeyondClassification}.
The emotion-aware audio large language model (EAA) integrates original features with features obtained through bidirectional acoustic and semantic interaction\cite{introDu2025EAA}.
AffectGPT uses multimodal prefusion, Modality-Specific Experts and Attention Reallocation (MoSEAR) combines modality experts with attention reallocation, and Omni-Perception Policy Optimization (OPPO) introduces perception rewards and modality-specific training\cite{introLian2025AffectGPT,introHan2025MoSEAR,introHan2026OPPO}.
These approaches motivate a complementary architectural question about how visually conditioned acoustic interaction relates to acoustic representations organized along their own timeline.

We propose DualAffect, a dual-path architecture organized around visual context and acoustic time.
Both pathways reuse the same WavLM sequence.
Visual-query fusion associates acoustic features with visual content, while the direct pathway independently projects summaries of successive acoustic intervals into ordered decoder tokens.
The shared decoder thus accesses joint audiovisual context and temporal vocal evidence through two organizations of the same encoding.
Mixed-task adaptation with task-dependent preservation aligns these inputs with affective and general answering behavior.
A single autoregressive model then generates emotion labels, short answers, or answer options according to the task instruction.

DualAffect achieves an average metric of 67.89 on Conflict-Aware Multimodal Emotion Reasoning (CA-MER) and accuracies of 62.88\% on the EmotionTalk test set and 49.87\% on Daily-Omni.
With model parameters fixed, zeroing the direct tokens reduces EmotionTalk validation accuracy by 9.75 percentage points and the CA-MER average metric by 2.66 points.
The largest CA-MER subset effect occurs when the overall emotion agrees with audio but differs from vision.

Our main contributions are threefold.
\begin{itemize}
  \item We propose dual-path paralinguistic integration for generative emotion understanding. DualAffect combines visually conditioned acoustic interaction with direct temporal acoustic inputs, organizing shared features around visual context and acoustic time within a single language decoder.
  \item We reveal the role of acoustic evidence organization in generative emotion understanding. Pathway interventions show that direct temporal integration complements visually conditioned fusion in the trained model, with its strongest effect in audio-aligned contexts.
  \item DualAffect achieves an average metric of 67.89 on CA-MER and accuracies of 62.88\% on the EmotionTalk test set and 49.87\% on Daily-Omni, demonstrating unified emotion recognition and general audiovisual question answering.
\end{itemize}

\section{Related Work}
\label{sec:related_work}

\subsection{Multimodal Emotion Understanding}

Multimodal emotion understanding combines affective cues with their contextual dependencies.
L-GrIN learns graph representations of affective sequences\cite{relatedShirian2022LGrIN}, while SDT combines intra- and intermodal Transformers with gated fusion and self-distillation\cite{relatedMa2024SDT}.
IMDNet and HyperCRM model conversational dependencies through speaker and modality attributes and hypergraph relations\cite{relatedShi2025IMDNet,relatedLu2025HyperCRM}.
Structured information propagation also appears in video segmentation through memory sampling, spatiotemporal propagation, dynamic routing, and representation decomposition\cite{relatedDang2023IMSFR,relatedDang2024STCM,relatedDang2025ASM,relatedDang2024TSPSM,relatedDang2024STDR,relatedDang2026FastTrack,relatedDang2026VDN,relatedDang2025DeSa2VA}.
For affective modeling, the central lesson is to organize temporal and cross-modal dependencies alongside the underlying perceptual features.

Complementary information is also a central concern in multimodal fusion.
Multimodal information bottleneck methods constrain unimodal and fused representations to retain compact, prediction-relevant features\cite{relatedMai2023MIB}.
SIMSUF combines dominant-modality selection with shared and specific representation enhancement\cite{relatedHuang2024SIMSUF}.
Lin and Hu align modality spaces through hybrid-modal attention and dynamic representation adjustment\cite{relatedLin2024Shifting}, while ROSA addresses uncertain missing modalities\cite{relatedLi2025ROSA}.
These discriminative approaches provide foundations for learning and combining affective representations.
Generative models extend this problem to the interface through which those representations condition language outputs.

Generative affective models express emotion judgments, descriptions, and explanations through a shared language interface.
Emotion-LLaMA jointly learns recognition and reasoning from multimodal affective instructions\cite{introCheng2024EmotionLLaMA}.
AffectGPT supports open-vocabulary descriptions through multimodal prefusion and affective training data\cite{introLian2025AffectGPT}, while Omni-Emotion augments a video language model with facial and acoustic perception\cite{relatedYang2025OmniEmotion}.
These methods emphasize affective supervision and specialized representations.
DualAffect examines their integration into generation, combining visually conditioned acoustic features with temporally organized acoustic tokens to expose both cross-modal context and ordered vocal cues to the decoder.

\subsection{Paralinguistic Representations and Language Model Integration}

Speech language models map acoustic representations into the input space of a language decoder.
SALMONN combines Whisper and BEATs features through a window-level Q-Former\cite{relatedTang2024SALMONN}, and WavLLM integrates complementary Whisper and WavLM representations\cite{introHu2024WavLLM}.
BLSP-Emo extends semantic alignment with emotion-aware training\cite{relatedWang2024BLSPEmo}.
Kang et al. show that adapting speech encoders and interface modules can make paralinguistic information influence responses from a frozen language model\cite{relatedKang2025Frozen}.
Together, these studies establish acoustic adaptation as an effective means of connecting nonlexical speech cues to language generation.

More closely related work studies the interaction and organization of affective features.
Beyond Classification combines speech and emotion-specific representations for multitask question answering and affective reasoning\cite{introZhang2025BeyondClassification}.
EAA retains original and bidirectionally interacted semantic and acoustic features as language model inputs\cite{introDu2025EAA}.
Its interactions operate between audio representations.
DualAffect instead combines visual-query fusion with direct temporal integration of the same acoustic sequence.
This distinction allows us to examine the role of acoustic evidence indexed by time alongside acoustic evidence organized by visual queries.

\subsection{Evidence Utilization in Affective Generation}

Evidence utilization involves both learning complementary cues and connecting them to emotion judgments.
HKD-MER transfers feature- and label-level knowledge from dominant to nondominant modalities to address learning shortcuts\cite{relatedSun2024HKDMER}.
EMERSK combines facial, postural, gait, and background information with situational knowledge to explain emotion predictions\cite{relatedPalash2024EMERSK}.
These approaches address complementary aspects of evidence use, the distribution of learning across modalities and the connection between predictions and interpretable observations.
Generative affective models additionally require the learned cues to influence the decoder's answer distribution.

Recent diagnostic studies distinguish encoded information from its use in generation.
LISTEN isolates lexical and acoustic emotion cues to reveal lexical dependence in audio language models\cite{introChen2026LISTEN}.
VoxParadox connects layer-wise paralinguistic probing to response behavior and introduces prompt-conditioned layer mixing and preference optimization\cite{introPang2026VoxParadox}.
Heard but Not Heeded similarly distinguishes encoded expressive style from output sensitivity to that style\cite{relatedKoduru2026Heard}.
These findings motivate studying both representation quality and the routes through which representations reach the decoder.
Our pathway analysis targets the latter by intervening on direct acoustic inputs while retaining audiovisual fusion.

Multimodal affective reasoning also depends on modality preferences.
MoSEAR uses modality experts and attention reallocation to address audio bias under emotional conflict\cite{introHan2025MoSEAR}, while EmoMM studies conflict-aware attention-head interventions\cite{relatedSun2026EmoMM}.
OPPO introduces perception rewards and modality-specific training\cite{introHan2026OPPO}.
AffectOmni combines people-focused and temporal-order rewards with executable evidence instructions for visual verification\cite{wang2026affectomni}.
These approaches emphasize evidence supervision and grounding.
DualAffect focuses on the internal interface to generation, supplying visually conditioned and temporally organized acoustic inputs and measuring how direct access affects the final answer.

\section{Method}
\label{sec:method}

\begin{figure*}[t]
  \centering
  \includegraphics[width=\textwidth,pagebox=cropbox]{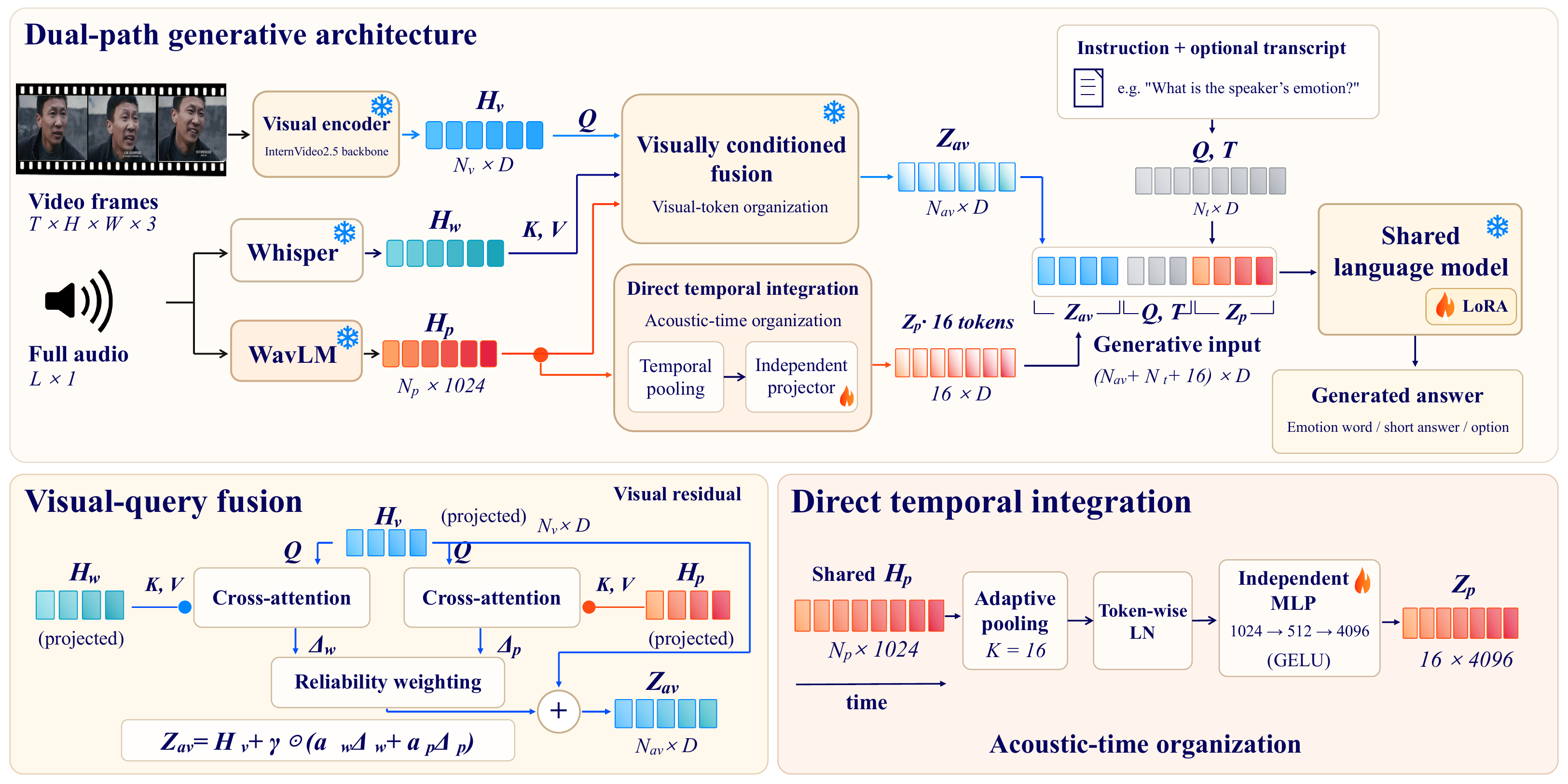}
  \caption{Overview of DualAffect. The upper panel shows two pathways that share WavLM features and condition one language decoder. Visual-query fusion integrates Whisper and WavLM features into audiovisual context. The direct pathway compresses the WavLM sequence into 16 ordered tokens through an independent projector. The lower panels detail the two pathways. Instructions and available transcripts enter as text embeddings, and the shared decoder generates the answer.}
  \label{fig:architecture}
\end{figure*}

DualAffect maps video, audio, and a task instruction to an answer through a shared language decoder.
As shown in Fig.~\ref{fig:architecture}, it extends an existing audiovisual fusion pathway with direct temporal acoustic inputs.
The fusion pathway uses visual queries to construct audiovisual context, while the direct pathway compresses the shared acoustic sequence into ordered tokens.
The resulting inputs organize acoustic evidence around visual content and acoustic time, respectively.
We describe their construction and integration before presenting the joint adaptation objective.

\subsection{Problem Formulation and Overview}
\label{sec:method_overview}

Given a video $V$, audio $A$, and task instruction $Q$, the model generates an answer $Y=(y_1,\ldots,y_L)$ of length $L$.
An available transcript is included as $T$, which is empty otherwise.
The instruction specifies the task and answer format, allowing emotion recognition and audiovisual question answering to share the same interface.
Training examples additionally provide a target answer $Y^*$.
The model learns to predict this sequence from the audiovisual inputs, transcript, and instruction through conditional language modeling.

InternVideo2.5 supplies the video language backbone\cite{methodWang2025InternVideo25}, while Whisper and WavLM extract complementary acoustic features\cite{methodRadford2023Whisper,methodChen2022WavLM}.
We denote the visual tokens by $H_{\mathrm v}$ and the acoustic sequences by $H_{\mathrm w}$ and $H_{\mathrm p}$.
The retained fusion pathway combines these features into audiovisual context $Z_{\mathrm{av}}$.
Our direct pathway compresses and independently projects $H_{\mathrm p}$ into paralinguistic tokens $Z_{\mathrm p}$.
Both inputs condition autoregressive generation through the same decoder:
\begin{equation}
  \begin{aligned}
    &p_{\theta}(Y\mid V,A,Q,T)\\
    &\quad=\prod_{t=1}^{L}p_{\theta}
    (y_t\mid y_{<t},Z_{\mathrm{av}},Z_{\mathrm p},Q,T),
  \end{aligned}
  \label{eq:method_generation}
\end{equation}
where $\theta$ denotes the model parameters and $y_{<t}$ is the generated answer prefix.
Each answer token thus depends on the joint audiovisual context, direct acoustic tokens, and textual instruction within one decoding process.

\subsection{Visually Conditioned Acoustic Interaction}
\label{sec:method_context}

The fusion pathway associates acoustic features with visual content through cross-attention.
The video backbone produces $H_{\mathrm v}\in\mathbb{R}^{N_{\mathrm v}\times d}$, where $N_{\mathrm v}$ is the visual token count and $d$ is the decoder embedding dimension.
Whisper and WavLM produce sequences of lengths $N_{\mathrm w}$ and $N_{\mathrm p}$, mapped to dimension $d$ by projectors $P_{\mathrm w}$ and $P_{\mathrm p}$.
We retain features according to the valid audio length and normalize WavLM features at each time position.
Long audio is encoded in chronological chunks, whose features and absolute timestamps are concatenated.

Two independent attention modules use visual tokens as queries and acoustic tokens as keys and values to compute audio-conditioned visual updates:
\begin{equation}
  \begin{aligned}
    \Delta_{\mathrm w}
      &=\mathcal{A}_{\mathrm w}(H_{\mathrm v},P_{\mathrm w}(H_{\mathrm w})),\\
    \Delta_{\mathrm p}
      &=\mathcal{A}_{\mathrm p}(H_{\mathrm v},P_{\mathrm p}(H_{\mathrm p})).
  \end{aligned}
  \label{eq:method_cross_attention}
\end{equation}
Here, $\mathcal{A}_{\mathrm w}$ and $\mathcal{A}_{\mathrm p}$ include input normalization, temporal information, and output projection, with $\Delta_{\mathrm w},\Delta_{\mathrm p}\in\mathbb{R}^{N_{\mathrm v}\times d}$.
The two updates retain the visual token layout while incorporating features from their respective acoustic encoders.
To weight these updates, the fusion module averages each over tokens and linearly maps the result to evidence scores $\ell_m\in\mathbb{R}^{C}$.
Here, $m\in\{\mathrm w,\mathrm p\}$ and $C=6$ is the module's internal evidence dimension.
Nonnegative evidence determines stream reliability, which is normalized across the two streams:
\begin{equation}
  \begin{aligned}
    r_m&=1-\frac{C}{C+\sum_{c=1}^{C}\operatorname{softplus}(\ell_{m,c})},\\
    \alpha_m&=\frac{\exp(r_m)}{\exp(r_{\mathrm w})+\exp(r_{\mathrm p})}.
  \end{aligned}
  \label{eq:method_reliability}
\end{equation}
The weighted acoustic updates are added to the visual representation to form audiovisual context:
\begin{equation}
  Z_{\mathrm{av}}
  =H_{\mathrm v}+\gamma\odot
  (\alpha_{\mathrm w}\Delta_{\mathrm w}+\alpha_{\mathrm p}\Delta_{\mathrm p}),
  \label{eq:method_fusion}
\end{equation}
where $\gamma\in\mathbb{R}^{d}$ contains channel-wise scaling parameters and $\odot$ denotes element-wise multiplication broadcast over the token dimension.
This residual fusion retains the visual token structure while incorporating acoustic context.
The direct pathway described next complements this representation with inputs indexed by acoustic time rather than visual queries.

\subsection{Direct Integration of Temporal Acoustic Representations}
\label{sec:method_direct}

The direct pathway gives the decoder acoustic evidence organized along the original sound sequence.
It converts valid WavLM features into a fixed number of ordered tokens, retaining coarse temporal variation with a constant input length.
Let $H_{\mathrm p}=(h_0,\ldots,h_{N_{\mathrm p}-1})$ be the normalized valid sequence, with $h_i\in\mathbb{R}^{d_{\mathrm p}}$.
For output position $k\in\{0,\ldots,K-1\}$, define $s_k=\lfloor kN_{\mathrm p}/K\rfloor$ and $e_k=\lceil(k+1)N_{\mathrm p}/K\rceil$.
Adaptive average pooling then produces $K$ temporal summaries:
\begin{equation}
  \bar h_k=\frac{1}{e_k-s_k}\sum_{i=s_k}^{e_k-1}h_i,
  \qquad
  \bar H_{\mathrm p}=(\bar h_0,\ldots,\bar h_{K-1}).
  \label{eq:method_temporal_pooling}
\end{equation}
Pooling windows adapt to the valid sequence length and preserve chronological output order.
For chunked audio, pooling operates on the concatenated valid features from the complete clip.
The resulting sequence provides separate summaries of successive intervals rather than a single global average.
A two-layer projector with layer normalization (LN) and a Gaussian error linear unit (GELU) maps each summary into the decoder embedding space.
The projector is shared across temporal positions and independent of the fusion projectors:
\begin{equation}
  z_k=\operatorname{Linear}_{2}\!\left(
  \operatorname{GELU}\!\left(
  \operatorname{Linear}_{1}(\operatorname{LN}(\bar h_k))
  \right)\right),
  \label{eq:method_direct_projection}
\end{equation}
Here, LN operates token-wise, and both linear layers include biases.
We use $d_{\mathrm p}=1024$, a hidden dimension of $512$, output dimension $d=4096$, and $K=16$.
The resulting sequence $Z_{\mathrm p}=(z_0,\ldots,z_{K-1})\in\mathbb{R}^{K\times d}$ supplies a compact temporal acoustic input to the decoder.
Its independent projector learns the mapping for direct generative use while sharing the acoustic encoder with the fusion pathway.

During input construction, $Z_{\mathrm{av}}$ occupies the video embedding positions.
We append $Z_{\mathrm p}$ as consecutive chronological embeddings with a short description identifying them as acoustic inputs.
Instructions and available transcripts use standard text embeddings.
Visual queries determine the acoustic content of fused tokens, whereas acoustic intervals determine each direct token independently of vision.
Both organizations condition the same autoregressive answer, allowing the decoder to access shared acoustic features through visual context and temporal summaries.
Sections~\ref{sec:exp_direct} and~\ref{sec:exp_context} examine their use through pathway intervention and modality-relationship analysis.

\subsection{Joint Adaptation for Unified Generation}
\label{sec:method_training}

We train emotion recognition and audiovisual question answering through instruction-conditioned generation.
Each example supplies audiovisual inputs, an instruction, an optional transcript, and a target answer.
Instructions and targets determine the answer format, while the acoustic modules, decoder, and output vocabulary remain shared.
Let $\mathcal{C}_x=(Z_{\mathrm{av}},Z_{\mathrm p},Q,T)$ denote the input conditions and $L_x$ the target length, including the end marker.
The generative loss supervises the target sequence:
\begin{equation}
  \mathcal{L}_{\mathrm{gen}}(x)
  =-\frac{1}{L_x}\sum_{t=1}^{L_x}
  \log p_{\theta}(y_t^*\mid y_{<t}^*,\mathcal{C}_x).
  \label{eq:method_generation_loss}
\end{equation}
Joint adaptation starts from a model with trained audiovisual fusion and a direct acoustic interface, then mixes affective data with general question-answering examples.
We freeze the perception encoders and existing fusion module and update the direct projector and language model low-rank adaptation (LoRA) parameters\cite{methodHu2022LoRA}.
The projector adapts the acoustic input mapping, while LoRA adjusts how the decoder uses direct acoustic tokens and joint context.
This parameter scope concentrates optimization on the interface between acoustic evidence and answer generation.

A frozen copy of the initial model provides a teacher distribution at selected answer positions.
Let $\Omega_x$ denote these positions for example $x$.
The teacher and student distributions, $p_{0,x,t}^{(\tau)}$ and $p_{\theta,x,t}^{(\tau)}$, are obtained by applying softmax to vocabulary logits scaled by temperature $\tau$.
Both models receive the same raw inputs and target prefix and construct their own continuous representations.
We preserve selected output distributions using the temperature-scaled Kullback-Leibler (KL) divergence:
\begin{equation}
  \mathcal{L}_{\mathrm{keep}}(x)
  =\frac{\tau^2}{|\Omega_x|}\sum_{t\in\Omega_x}
  D_{\mathrm{KL}}\!\left(
  p_{0,x,t}^{(\tau)}\,\Vert\,p_{\theta,x,t}^{(\tau)}
  \right).
  \label{eq:method_preservation}
\end{equation}
We set $\mathcal{L}_{\mathrm{keep}}(x)=0$ when $\Omega_x$ is empty.
Teacher parameters remain fixed throughout adaptation.

The preservation scope follows the task.
New-domain affective examples use target-answer supervision alone.
Affective replay constrains emotion-word positions, while general question answering constrains the complete target response.
This design assigns different output positions to new-domain learning, existing affective judgments, and general answering behavior within one training process.
Combining generative supervision with task-dependent preservation gives the final objective:
\begin{equation}
  \mathcal{L}
  =\mathbb{E}_{x\sim\mathcal{D}_{\mathrm{mix}}}
  \left[\mathcal{L}_{\mathrm{gen}}(x)
  +\lambda\mathcal{L}_{\mathrm{keep}}(x)\right],
  \label{eq:method_total_loss}
\end{equation}
where $\mathcal{D}_{\mathrm{mix}}$ is the mixed-task training distribution and $\lambda$ controls the preservation weight.

EmotionTalk supplies new-domain conversational emotion data, while Multimodal Emotion Recognition (MER) and Multimodal Emotion Recognition and Reasoning (MERR) provide affective replay.
General question-answering examples support broader audiovisual responses.
All sources update the same direct projector and decoder LoRA parameters, with data allocation detailed in supplementary Table~S1.
At inference, a single adapted model generates emotion labels or audiovisual answers according to the task instruction and~\eqref{eq:method_generation}.
The complete training and inference procedure is provided in supplementary Section~S1-B and Algorithm~S1.

\section{Experiments}
\label{sec:experiments}

Our experiments evaluate affective recognition, general audiovisual understanding, and the role of direct acoustic access.
CA-MER tests emotion judgments across modality relationships, EmotionTalk evaluates conversational emotion recognition, and Daily-Omni tests DualAffect's audiovisual question answering.
We then examine direct-pathway interventions and matched-training comparisons of preservation objectives, followed by context-dependent pathway effects and task-specific generation settings.
This organization connects system-level performance to the architectural and training choices introduced in Section~\ref{sec:method}.

\subsection{Experimental Setup}
\label{sec:exp_setup}

\subsubsection{Datasets and Evaluation Metrics}

CA-MER groups emotion judgments by modality-specific label relationships\cite{introHan2025MoSEAR}.
In Audio-aligned and Video-aligned examples, the overall label agrees with one modality and differs from the other.
In Consistent examples, all three labels agree.
We evaluate 1,483 available examples, comprising 489, 498, and 496 examples in these subsets.
Our fixed-label evaluation reports accuracy (Acc), balanced accuracy (BAcc), computed as macro-averaged class recall, and their mean $\mathrm{Avg}_{\mathrm{cls}}=(\mathrm{Acc}+\mathrm{BAcc})/2$.
Published open-vocabulary results retain their source metric definitions, with the evaluation settings identified in Table~\ref{tab:camer_main}.

EmotionTalk provides Chinese conversational video, audio, and transcripts\cite{expSun2026EmotionTalk}.
We use seven-class overall emotion labels and the official split of 15,413 training, 1,908 validation, and 1,929 test examples.
Main results report official-test accuracy, supplemented by BAcc, weighted F1 (W-F1), and macro F1 (M-F1) in the analyses.
Checkpoint selection and training-objective comparisons use fixed development sets of 512 EmotionTalk and 259 MER examples.
The full EmotionTalk validation set supports adaptation and direct-pathway comparisons.

Daily-Omni evaluates audiovisual question answering across six categories, audiovisual event alignment, comparison, contextual understanding, event ordering, inference, and reasoning\cite{expZhou2025DailyOmni}.
We evaluate all 1,197 multiple-choice questions and report overall, category-level, and video-duration accuracy.
Duration groups contain questions associated with 30-second and 60-second videos.
Accuracy is aggregated by question count, with all examples in the corresponding group included in the denominator.
Across datasets, invalid answers count as errors and performance metrics are percentages.

\subsubsection{Implementation and Training Details}

DualAffect uses InternVideo2.5, Whisper-medium, and WavLM-Large initialized from an emotion diarization model.
The direct projector maps 1024-dimensional acoustic features through a 512-dimensional hidden layer to 16 tokens of dimension 4096.
We sample 8 frames for training and emotion evaluation and 16 frames for Daily-Omni, at a resolution of $448\times448$.
All evaluations use complete audio resampled to 16 kHz.
Audio longer than 30 seconds is encoded in chunks, and valid features are concatenated chronologically.

Final adaptation starts from a model with a trained direct acoustic interface and transcript-conditioned mixed-task adaptation.
Training combines new-domain emotion examples, affective replay, and general question answering according to supplementary Table~S1.
General questions come from the existing Video-R1 data.
We freeze the perception encoders and fusion module and train the direct projector and decoder LoRA parameters.
LoRA targets the joint query-key-value and attention output projections, using rank 32, scaling factor 64, and dropout 0.05.

We optimize with AdamW, using learning rates of $2\times10^{-6}$ for LoRA and $5\times10^{-5}$ for the projector, weight decay 0.01, and gradient accumulation over 8 examples.
Learning rates increase during a 20-update warmup and then remain constant.
Preservation uses $\lambda=1$ and $\tau=2$.
Within the 400-update adaptation run, checkpoint selection requires at least a 3-point EmotionTalk accuracy gain over initialization and at most a 2-point MER BAcc decrease on the development sets.
This rule selects update 200 as the final DualAffect model.

DualAffect uses a shared language decoder across all three evaluations.
CA-MER uses generation constrained to its five emotion labels, EmotionTalk uses short emotion answers, and Daily-Omni uses answer-option generation.
Decoding is greedy, with a maximum of 256 tokens for short answers.
The CA-MER limit follows the longest valid answer sequence.
Section~\ref{sec:exp_generation} compares unconstrained and five-label constrained generation on CA-MER.
Task prompts and answer extraction rules are provided in supplementary Section~S2.

\subsection{Emotion Understanding across Contexts and General Question Answering}
\label{sec:exp_main}

\subsubsection{Emotion Judgments under Different Modality Relationships}

CA-MER tests emotion judgments when visual and acoustic labels agree or diverge.
Table~\ref{tab:camer_main} presents published results alongside DualAffect across its three modality-relationship subsets.
The comparison includes audio, visual, and audiovisual models, ranging from general multimodal systems to affective models such as Emotion-LLaMA and MoSEAR.
With fixed-label generation, DualAffect achieves 66.49 accuracy, 69.30 BAcc, and an average metric of 67.89.
The subset structure further allows us to relate recognition performance to the evidence relationships within each clip.

\begin{table*}[t]
  \centering
  \caption{CA-MER results. Published scores from MoSEAR\cite{introHan2025MoSEAR} use open-vocabulary set-level accuracy and recall. DualAffect uses fixed-label generation on 1,483 examples, with Rec. denoting macro-averaged class recall and Avg denoting $\mathrm{Avg}_{\mathrm{cls}}$. Across the main tables, bold marks the best listed value and blue shading highlights our best values.}
  \label{tab:camer_main}
  \setlength{\tabcolsep}{3pt}
  \resizebox{\textwidth}{!}{%
  \begin{tabular}{lrrrrrrrrrrrr}
    \toprule
    \multirow{2}{*}{Method}
      & \multicolumn{3}{c}{Audio-aligned}
      & \multicolumn{3}{c}{Video-aligned}
      & \multicolumn{3}{c}{Consistent}
      & \multicolumn{3}{c}{Overall} \\
    \cmidrule(lr){2-4}\cmidrule(lr){5-7}\cmidrule(lr){8-10}\cmidrule(lr){11-13}
    & Acc & Rec. & Avg & Acc & Rec. & Avg & Acc & Rec. & Avg & Acc & Rec. & Avg \\
    \midrule
    SALMONN~\cite{relatedTang2024SALMONN} & 53.35 & 34.19 & 43.77 & 29.37 & 22.56 & 25.97 & 47.72 & 38.53 & 43.13 & 43.48 & 31.76 & 37.62 \\
    mPLUG-Owl3~\cite{expYe2025MPLUGOwl3} & 41.08 & 41.01 & 41.04 & 40.71 & 42.01 & 41.36 & 49.29 & 39.08 & 44.18 & 43.69 & 40.70 & 42.19 \\
    Chat-UniVi~\cite{expJin2024ChatUniVi} & 42.28 & 36.31 & 39.30 & 43.43 & 42.42 & 42.92 & 50.38 & 45.95 & 48.16 & 45.36 & 41.56 & 43.46 \\
    VITA-1.5~\cite{expFu2025VITA15} & 57.51 & 48.15 & 52.83 & 49.36 & 46.60 & 47.98 & 57.17 & 52.52 & 54.84 & 54.68 & 49.09 & 51.88 \\
    Emotion-LLaMA~\cite{introCheng2024EmotionLLaMA} & 59.94 & 51.63 & 55.78 & 47.66 & 51.45 & 49.56 & 59.23 & 63.80 & 61.52 & 55.61 & 55.63 & 55.62 \\
    MoSEAR~\cite{introHan2025MoSEAR} & 65.33 & 55.26 & 60.30 & \textbf{58.42} & 54.28 & 56.35 & 67.68 & 65.87 & 66.77 & 63.81 & 58.47 & 61.14 \\
    \midrule
    DualAffect (Ours) & \ourbest{68.30} & \ourbest{63.80} & \ourbest{66.05} & 55.42 & \ourbest{62.83} & \ourbest{59.13} & \ourbest{75.81} & \ourbest{78.35} & \ourbest{77.08} & \ourbest{66.49} & \ourbest{69.30} & \ourbest{67.89} \\
    \bottomrule
  \end{tabular}}
\end{table*}

DualAffect achieves average metrics of 77.08 on Consistent, 66.05 on Audio-aligned, and 59.13 on Video-aligned.
The highest value occurs when visual, acoustic, and overall labels agree.
The two conflict subsets reveal distinct behavior according to which modality matches the overall judgment.
These results provide the context for the pathway analysis in Section~\ref{sec:exp_context}, which measures how direct acoustic access affects each subset while keeping the model and remaining inputs fixed.

\subsubsection{Generative Emotion Recognition on EmotionTalk}

EmotionTalk evaluates seven-class conversational emotion recognition from video, audio, and transcripts.
DualAffect produces emotion answers through the same decoder used for the other tasks.
Table~\ref{tab:emotiontalk_main} compares it with the Multimodal Cyclic Translation Network (MCTN) and Multimodal Factorization Model (MFM) from the initial dataset paper.
These baselines use HuBERT-Base, Baichuan-7B, and CLIP-Large representations, whereas DualAffect predicts answer tokens from its audiovisual inputs.
The comparison places the generative system alongside established feature-fusion approaches on the official test set.

\begin{table}[t]
  \centering
  \caption{EmotionTalk accuracy for selected feature-fusion baselines and DualAffect. External scores use the HuBERT-Base configuration from Table 3 of the initial dataset paper\cite{expSun2025EmotionTalk}. Four covers happy, angry, sad, and neutral, while All covers seven classes. The omitted entry is unevaluated.}
  \label{tab:emotiontalk_main}
  \small
  \setlength{\tabcolsep}{3pt}
  \begin{tabular}{llrr}
    \toprule
    Method & Fusion or Output & Four & All \\
    \midrule
    MCTN~\cite{expPham2019MCTN} & Frame-level & 65.34 & 47.80 \\
    MFM~\cite{expTsai2019MFM} & Frame-level & \textbf{75.94} & 59.51 \\
    \midrule
    DualAffect (Ours) & Token generation & $\cdots$ & \ourbest{62.88} \\
    \bottomrule
  \end{tabular}
\end{table}

DualAffect achieves 62.88 test accuracy, exceeding the listed MCTN and MFM results by 15.08 and 3.37 percentage points, respectively.
DualAffect also obtains 49.70 BAcc, 60.94 W-F1, and 49.61 M-F1.

To examine the change induced by adaptation, we also compare the initialization and final model on the full EmotionTalk validation set.
Both use the same architecture, media inputs, prompts, and deterministic answer extraction.
Accuracy increases from 53.46\% to 62.42\%, a gain of 8.96 percentage points.
BAcc increases from 41.08 to 48.60, while M-F1 and W-F1 rise from 41.18 to 49.66 and from 53.47 to 61.23, respectively.
Invalid answers decrease from 174 to 39, indicating improved adherence to the task vocabulary through the shared decoder.
The complete adaptation comparison is provided in supplementary Section~S1-D.

\subsubsection{General Audiovisual Question Answering with a Unified Model}

Daily-Omni tests whether DualAffect supports general audiovisual understanding.
Table~\ref{tab:daily_main} compares representative open-source models across six question categories and two video durations.
DualAffect answers 597 questions correctly for 49.87 overall accuracy, exceeding the listed VideoLLaMA2 and Unified-IO-2 results and approaching Ola's 50.71.
These answers use the same model and decoder as the emotion evaluations, with the task instruction specifying the required answer option.

\begin{table*}[t]
  \centering
  \caption{Daily-Omni accuracy of representative open-source audiovisual models. External scores follow the official leaderboard\cite{expDailyOmniLeaderboard}, accessed October 10, 2026. DualAffect uses 16 frames and complete audio. Overall accuracy is aggregated over all questions.}
  \label{tab:daily_main}
  \setlength{\tabcolsep}{3.5pt}
  \resizebox{\textwidth}{!}{%
  \begin{tabular}{lrrrrrrrrr}
    \toprule
    Method & Alignment & Comparison & Context & Ordering & Inference & Reasoning & 30 s & 60 s & Overall \\
    \midrule
    Unified-IO-2 L~\cite{expLu2024UnifiedIO2} & 27.31 & 22.90 & 26.42 & 27.78 & 29.87 & 29.14 & 27.67 & 27.09 & 27.40 \\
    Unified-IO-2 XXL~\cite{expLu2024UnifiedIO2} & 25.63 & 31.30 & 26.42 & 25.82 & 35.06 & 29.71 & 26.74 & 30.00 & 28.24 \\
    Unified-IO-2 XL~\cite{expLu2024UnifiedIO2} & 30.25 & 30.53 & 25.39 & 29.08 & 33.12 & 21.71 & 28.13 & 28.55 & 28.32 \\
    VideoLLaMA2~\cite{expCheng2024VideoLLaMA2} & 35.71 & 35.88 & 35.75 & 31.70 & 40.91 & 34.29 & 38.02 & 31.82 & 35.17 \\
    Ola~\cite{expLiu2025Ola} & 40.34 & \textbf{61.07} & 40.41 & 43.46 & \textbf{63.64} & \textbf{69.71} & \textbf{51.47} & \textbf{49.82} & \textbf{50.71} \\
    \midrule
    DualAffect (Ours) & \ourbest{45.80} & 51.15 & \ourbest{46.11} & \ourbest{46.08} & 61.04 & 55.43 & 50.70 & 48.91 & 49.87 \\
    \bottomrule
  \end{tabular}}
\end{table*}

DualAffect achieves the highest listed scores in audiovisual event alignment, contextual understanding, and event ordering, with accuracies of 45.80, 46.11, and 46.08, respectively.
It obtains 50.70 on questions associated with 30-second videos and 48.91 on those associated with 60-second videos.
These results demonstrate general audiovisual question-answering capability after affective adaptation.
Together with CA-MER and EmotionTalk, they establish the cross-task coverage of DualAffect through a shared generative interface.

\begin{figure*}[t]
  \centering
  \includegraphics[width=\textwidth,pagebox=cropbox]{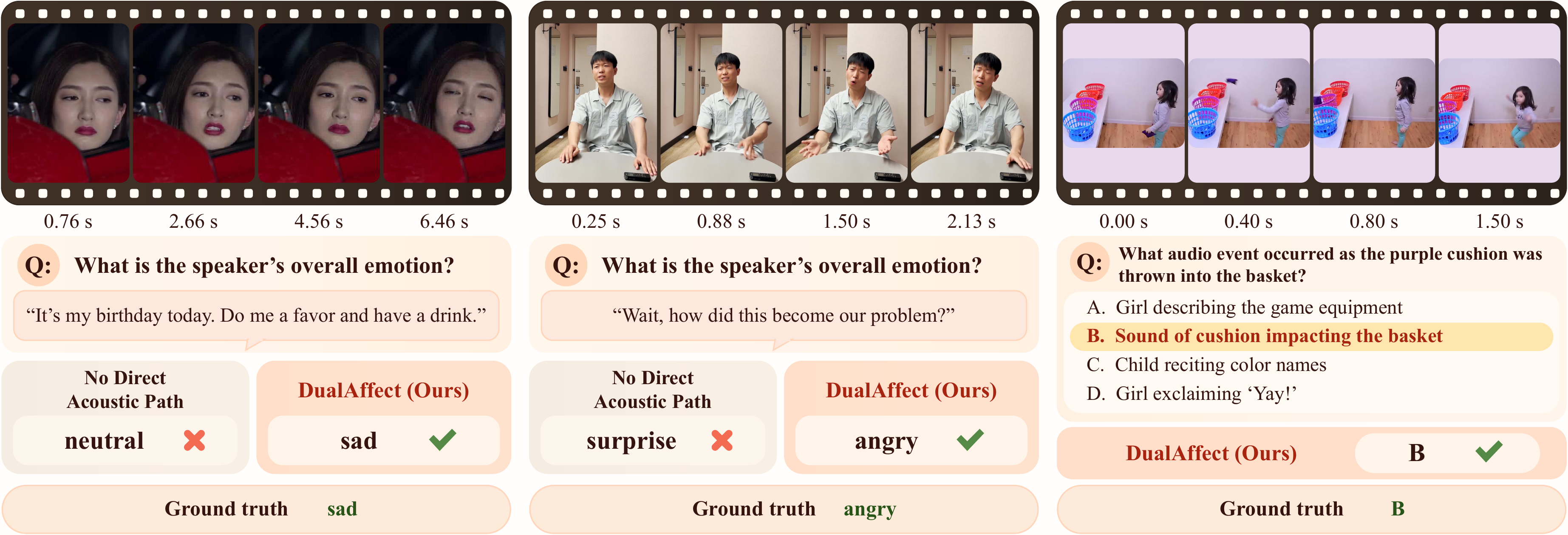}
  \caption{Qualitative results on CA-MER, EmotionTalk, and Daily-Omni, from left to right. No Direct Acoustic Path denotes the token-zeroing intervention with model parameters fixed in Section~\ref{sec:exp_direct}. The emotion examples show its effect on the answer, while the Daily-Omni example illustrates audiovisual event alignment through answer-option generation.}
  \label{fig:qualitative}
\end{figure*}

Fig.~\ref{fig:qualitative} illustrates how direct acoustic access affects individual emotion judgments.
In the CA-MER example, the birthday invitation does not explicitly express sadness, yet DualAffect predicts sad, matching the reference label.
Zeroing the direct acoustic tokens changes the prediction to neutral.
The EmotionTalk example distinguishes a change in emotion selection from a change in answer format.
The complete and intervened models generate angry and surprise, respectively, both valid emotion labels, but only the former matches the reference.
These paired examples show that direct acoustic input participates in selecting the final emotion even when visual-context fusion is retained, rather than merely affecting output formatting.

The Daily-Omni example shows the same model interpreting an audiovisual event.
Asked which sound occurs as a cushion is thrown into a basket, DualAffect correctly selects the sound of the cushion impacting the basket, associating the visible action with the corresponding acoustic event.
Together, the examples illustrate how one decoder produces emotion judgments and answers concrete audiovisual questions according to the task instruction.

\subsection{Ablation Studies}
\label{sec:exp_ablation}

We examine the architecture and training objective through two complementary comparisons.
The first measures how direct acoustic inputs affect the trained model while retaining audiovisual fusion.
The second varies the preservation objective during adaptation under matched initialization, data order, and training budget.
These comparisons address distinct questions about the model, how it uses the direct pathway at inference and how task-dependent constraints shape adaptation.

\subsubsection{Role of the Direct Paralinguistic Pathway}
\label{sec:exp_direct}

We test whether the final model uses direct acoustic evidence alongside the retained fusion pathway.
All DualAffect parameters are fixed, and the 16 direct paralinguistic tokens are replaced with zero vectors at inference.
Input length, video, transcripts, acoustic encodings, and fusion outputs remain unchanged.
The comparison therefore isolates the effect of information carried by the direct tokens within the trained model.
Table~\ref{tab:direct_route} reports results on CA-MER and the full EmotionTalk validation set.

\begin{table}[!htbp]
  \centering
  \caption{Direct-pathway intervention with DualAffect parameters fixed. Zeroed replaces the direct acoustic tokens with zero vectors, with audiovisual fusion unchanged. CA-MER uses fixed-label generation, and EmotionTalk uses the full validation set. Invalid counts invalid answers.}
  \label{tab:direct_route}
  \small
  \setlength{\tabcolsep}{3pt}
  \begin{tabular*}{\columnwidth}{@{\extracolsep{\fill}}llrc@{}}
    \toprule
    Dataset & Metric & Zeroed & DualAffect (Ours) \\
    \midrule
    \multirow{3}{*}{CA-MER}
      & Acc & 63.45 & \ourbest{66.49} \\
      & BAcc & 67.02 & \ourbest{69.30} \\
      & $\mathrm{Avg}_{\mathrm{cls}}$ & 65.23 & \ourbest{67.89} \\
    \midrule
    \multirow{4}{*}{EmotionTalk}
      & Acc & 52.67 & \ourbest{62.42} \\
      & BAcc & 40.73 & \ourbest{48.60} \\
      & M-F1 & 40.58 & \ourbest{49.66} \\
      & Invalid & 187 & \ourbest{39} \\
    \bottomrule
  \end{tabular*}
\end{table}

Direct-token zeroing reduces EmotionTalk accuracy by 9.75 percentage points, BAcc by 7.87 points, and M-F1 by 9.08 points.
Invalid answers increase from 39 to 187, connecting the pathway to both emotion prediction and answer validity.
On CA-MER, all answers follow the constrained label format, yet zeroing reduces accuracy by 3.03 points and the average metric by 2.66 points.
The consistent direction across datasets shows that the trained decoder uses direct acoustic evidence even when audiovisual fusion remains available.

The aggregate effects motivate a closer examination of the evidence relationships within CA-MER.
Section~\ref{sec:exp_context} separates Audio-aligned, Video-aligned, and Consistent examples to identify the contexts in which the final decoder is most sensitive to direct acoustic input.

\subsubsection{Role of the Joint Adaptation Strategy}
\label{sec:exp_preservation}

We compare generative supervision with replay alone, full-sequence preservation on replay answers, and task-dependent preservation.
All settings share the initialization and first 200 updates of the training sequence, including sample order, data proportions, learning rates, warmup, and random seed.
New EmotionTalk examples receive the same generative supervision, while replay preservation varies.
Table~\ref{tab:preservation} reports performance on the fixed EmotionTalk and MER development sets, separating new-domain recognition from existing affective performance.

\begin{table}[!htbp]
  \centering
  \caption{Adaptation objectives on fixed EmotionTalk and MER development sets with shared initialization and 200 updates. Initial denotes the model before adaptation. Full-seq. and Ours use full-sequence and task-dependent preservation, respectively.}
  \label{tab:preservation}
  \small
  \setlength{\tabcolsep}{2pt}
  \begin{tabular*}{\columnwidth}{@{\extracolsep{\fill}}lrrrrrr@{}}
    \toprule
    \multirow{2}{*}{Setting} & \multicolumn{3}{c}{EmotionTalk}
      & \multicolumn{3}{c}{MER} \\
    \cmidrule(lr){2-4}\cmidrule(lr){5-7}
    & Acc & BAcc & M-F1 & Acc & BAcc & M-F1 \\
    \midrule
    Initial & 55.66 & 43.33 & 41.90 & 64.86 & \textbf{62.94} & \textbf{61.79} \\
    Replay only & \textbf{65.23} & \textbf{50.68} & \textbf{51.91} & 64.48 & 59.25 & 59.55 \\
    Full-seq. & 64.45 & 47.74 & 47.90 & 64.48 & 57.87 & 58.41 \\
    \midrule
    Ours & 62.70 & 48.01 & 47.69 & \ourbest{65.25} & 62.02 & 61.32 \\
    \bottomrule
  \end{tabular*}
\end{table}

Task-dependent preservation raises EmotionTalk accuracy by 7.03 percentage points over initialization and obtains 62.02 MER BAcc.
Its MER BAcc exceeds replay-only and full-sequence preservation by 2.78 and 4.15 points, respectively.
Replay-only training gives the highest new-domain accuracy, while task-dependent preservation yields the strongest MER results among the adapted models.
The comparison identifies preservation scope as an effective control over adaptation behavior, supporting emotion-word constraints for affective replay and complete-response constraints for general question answering.

\subsection{Context-Dependent Use of Direct Acoustic Evidence}
\label{sec:exp_context}

The overall intervention results establish that the trained decoder uses the direct pathway.
We next examine how this use varies with the relationship between the visual, audio, and overall emotion labels.
CA-MER provides three complementary contexts for this analysis.
Audio-aligned examples require a judgment consistent with audio despite a different visual label, Video-aligned examples exhibit the reverse relationship, and Consistent examples align both modalities with the overall label.
This grouping connects the intervention to the evidence relationships that motivate the dual-path interface.

Table~\ref{tab:route_context} reports the complete model and direct-token-zeroed results for each group.
We hold model parameters, media inputs, transcripts, fusion outputs, and sequence length fixed in every comparison.
Both settings use identical five-label constrained generation, so the group differences are measured under the same answer space.
The reported drop is the complete-model score minus the intervened score, with $\mathrm{Avg}_{\mathrm{cls}}$ used throughout.

\begin{table}[!htbp]
  \centering
  \caption{CA-MER subset results under direct-token zeroing. Scores are $\mathrm{Avg}_{\mathrm{cls}}$, and Drop is the full-model score minus the zeroed score.}
  \label{tab:route_context}
  \small
  \setlength{\tabcolsep}{3pt}
  \begin{tabular}{lrrrr}
    \toprule
    Subset & Samples & DualAffect (Ours) & Zeroed & Drop \\
    \midrule
    Audio-aligned & 489 & 66.05 & 61.42 & 4.63 \\
    Video-aligned & 498 & 59.13 & 58.19 & 0.94 \\
    Consistent & 496 & 77.08 & 74.43 & 2.65 \\
    Overall & 1,483 & 67.89 & 65.23 & 2.66 \\
    \bottomrule
  \end{tabular}
\end{table}

Audio-aligned examples show the largest decrease, from 66.05 to 61.42, or 4.63 percentage points.
The corresponding decrease on Video-aligned examples is 0.94 points, from 59.13 to 58.19.
The difference between these intervention effects is 3.69 points.
Thus, the strongest response occurs precisely where the overall judgment follows vocal evidence rather than the visual label.
This pattern supports the motivation for giving acoustic evidence a direct route to the decoder alongside its visually conditioned representation.

Consistent examples also benefit from direct acoustic access, with a decrease from 77.08 to 74.43 after zeroing.
The 2.65-point change shows that the pathway is used when the two modality labels agree as well as when they differ.
Direct temporal integration therefore contributes to the final model across more than one evidence configuration.
Taken together, the three subsets reveal a context-dependent pattern of acoustic utilization, with the strongest contribution in audio-aligned judgments and an additional contribution when vocal and visual evidence support the same label.

\subsection{Task-Specific Answer Generation}
\label{sec:exp_generation}

We compare unconstrained short answers with five-label constrained generation on all 1,483 CA-MER examples, holding model parameters and audiovisual inputs fixed.
Both settings use the shared autoregressive decoder.
For constrained generation, the prompt lists five emotion labels, and each decoding step admits only valid label continuations.
Generation ends at the closing emotion tag, with a token limit equal to the longest valid answer sequence plus two.

\begin{table}[!htbp]
  \centering
  \caption{CA-MER generation settings for DualAffect (Ours). Model parameters and audiovisual inputs are fixed.}
  \label{tab:main_generation}
  \small
  \setlength{\tabcolsep}{4pt}
  \begin{tabular}{lrrr}
    \toprule
    Generation & Acc & BAcc & $\mathrm{Avg}_{\mathrm{cls}}$ \\
    \midrule
    Unconstrained short answer & 63.72 & 66.02 & 64.87 \\
    Five-label constrained & \ourbest{66.49} & \ourbest{69.30} & \ourbest{67.89} \\
    \bottomrule
  \end{tabular}
\end{table}

Table~\ref{tab:main_generation} shows that five-label generation produces 41 additional correct answers and improves $\mathrm{Avg}_{\mathrm{cls}}$ by 3.02 percentage points.
The gain is reflected in both accuracy and class-balanced recall.
Acoustic access and answer-space specification act at different stages of prediction.
The direct pathway supplies evidence to the decoder, whereas constrained generation aligns its output with the task vocabulary.
The pathway interventions retain this answer space throughout, isolating changes in acoustic access within the chosen generation setting.

\section{Conclusion}
\label{sec:conclusion}

DualAffect connects paralinguistic perception to answer generation through two organizations of a shared acoustic sequence, visual-context fusion and direct temporal integration.
Pathway interventions show that the final model uses direct acoustic evidence alongside existing fusion, with effects on both emotion datasets.
The strongest CA-MER subset response occurs when the overall emotion follows audio rather than vision.
This finding links direct acoustic utilization to modality-specific evidence relationships and identifies representation organization as a practical design choice for generative affective understanding.
The adaptation comparison further shows that task-dependent preservation supports new-domain learning while maintaining stronger existing-task performance than the alternative adaptation objectives.
Together, these results establish a concrete route to affective enhancement through acoustic input design and targeted generative adaptation.
The observed variation across modality relationships motivates future study of context-dependent control over direct acoustic access.

\section*{Acknowledgments}
This work was supported in part by the National Natural Science Foundation of China under Grants 62606208, 62606209, 62227807, and U24B20186, by the National Science and Technology Major Project on Brain Science and Brain-like Intelligence Technology under Grants 2021ZD0200600 and 2021ZD0200408, and by the Talent Scientific Fund of Lanzhou University.
We also acknowledge support from the WQ \& UCAS Research Academy Intelligent Computing Center (WRA-ICC) and the Supercomputing Center of Lanzhou University.

\bibliographystyle{IEEEtran}
\bibliography{refs_phantom}

\end{document}